\documentclass[aps,prd,onecolumn,groupedaddress,nofootinbib,showpacs, showkeys,amssymb]{revtex4}
\usepackage{latexsym,float}
\usepackage{epsfig,graphics}
\usepackage{epstopdf}
\usepackage{amssymb}
\usepackage{verbatim}
\usepackage{multirow}
\usepackage{color}
\usepackage{tikz}
\usepackage[utf8]{inputenc}
\usetikzlibrary{matrix}

\def\gtwid{\mathrel{\raise.3ex\hbox{$>$\kern-.75em\lower1ex\hbox{$\sim$}}}}
\def\ltwid{\mathrel{\raise.3ex\hbox{$<$\kern-.75em\lower1ex\hbox{$\sim$}}}}
\def\square{\kern1pt\vbox{\hrule height 1.2pt\hbox{\vrule width 1.2pt\hskip 3pt
			\vbox{\vskip 6pt}\hskip 3pt\vrule width 0.6pt}\hrule height 0.6pt}\kern1pt}
\begin{document}

\title{Nonlocal Teleparallel  Cosmology}

\author{Sebastian Bahamonde}
\email{sebastian.beltran.14@ucl.ac.uk}
\affiliation{Department of Mathematics, University College London,
	Gower Street, London, WC1E 6BT, United Kingdom}

\author{Salvatore Capozziello}
\email{capozziello@na.infn.it}
\affiliation{Dipartimento di Fisica "E. Pancini", Universit\'a di Napoli
	\textquotedblleft{Federico II}\textquotedblright, Napoli, Italy,}
\affiliation{Gran Sasso Science Institute, Via F. Crispi 7, I-67100, L' Aquila,
	Italy,}
\affiliation{INFN Sez. di Napoli, Compl. Univ. di Monte S. Angelo, Edificio G, Via
	Cinthia, I-80126,
	Napoli, Italy.}
\author{Mir Faizal}
\email{mirfaizalmir@googlemail.com}
\affiliation{Irving K. Barber School of Arts and Sciences, University of British
	Columbia - Okanagan, 3333 University Way, Kelowna,   British Columbia V1V 1V7, Canada.}
\affiliation{ Department of Physics and Astronomy, University of Lethbridge,
	Lethbridge, Alberta, T1K 3M4, Canada}
	
\author{Rafael C. Nunes}
\email{rcnunes@fisica.ufjf.br}
\affiliation{Departamento de F\'isica, Universidade Federal de Juiz de Fora, 36036-330, Juiz de Fora, MG, Brazil}

\begin{abstract}
Even though it is not possible to  differentiate General Relativity from Teleparallel Gravity using  classical experiments, it could be  possible to discriminate between them by quantum gravitational effects.
These effects have motivated the introduction of  nonlocal deformations of General Relativity, and similar effects are also expected 
to occur in Teleparallel Gravity.   Here, we study    nonlocal deformations of  Teleparallel Gravity along with  its cosmological solutions.   
We observe  that  Nonlocal  Teleparallel Gravity (like nonlocal General Relativity) is consistent with  the present cosmological data obtained 
by SNe Ia + BAO + CC + $H_0$ observations.  Along this track,  future experiments  probing  nonlocal effects  could  be used to test  
whether General Relativity or Teleparallel Gravity give the most consistent picture of gravitational interaction. 
\end{abstract}
\date{\today}

\maketitle
	
\section{Introduction}
General Relativity (GR) tells us that gravitational interaction is described by the  curvature of  torsion-less spacetimes. On the other hand, it is  possible to    describe  gravity by the torsion of spacetime, so that the curvature picture is not necessary. A theory  where gravity is described by the torsion of spacetime (without curvature) is called the Teleparallel Equivalent of General Relativity (TEGR) \cite{tele,  tele1, ft12, ft14}.  
Even though  these two approaches are fundamentally different, they produce the same classical field equations. Thus, both theories predict the same dynamics for 
classical gravitational systems, and so classical gravitational experiments cannot be used to test  which of them gives the  correct theory of gravity. In other words, they are equivalent at classical level.
	
However, because these theories are conceptually different, they are expected to  produce  different quantum effects. A important remark is in order at this point. We can deal with TEGR only at classical level because it produces the same classical field equations as GR. Considering quantum effects and nonlocality, it is improper to speak of equivalence of the two theories since they could be fundamentally different. Due to this fact, we will speak of Teleparallel Gravity in general and of TEGR in the classical case.  

Even though we do not have a fully developed quantum theory of gravity, there are various proposals for quantum gravity, and a universal prediction from almost all of these approaches seem to be the existence of an intrinsic extended structure in the geometry of spacetime \cite{univ4, univ5}, and such an extended structure would be related to  an effective nonlocal behavior for spacetime \cite{univ1, univ2, modesto1,modesto2}.  For example,  in perturbative string theory,
it is not possible to measure spacetime below string length scale, 
as the string is the smallest  available probe. As  it is not possible define point-like local structures, 
string theory produces an effective nonlocal behavior \cite{st1, st2}. Similarly, there  is an intrinsic minimal  area in loop-quantum gravity   \cite{loop}, and this   extended structure   is expected to produce a  nonlocal behavior.  It can    be argued, from black hole physics, that any theory 
of quantum gravity should present intrinsic extended structures of the order of the Planck length, 
and it would not be possible to probe the spacetime below this scale. 
In fact,  the energy needed to probe the spacetime below this scale is more than 
the energy needed to form a mini black hole in that region of spacetime \cite{z4,z5}.

Thus, quantum gravitational effects produce  effective extended structures in spacetime that  would lead to   nonlocality \cite{univ4, univ5}. Hence,  
it can be   argued  that the first order corrections from quantum gravity will
produce nonlocal deformations of GR  \cite{jm, jm12,Elizalde:1995tx}, and this will, in turn, produce a  nonlocality in  cosmology. 
The effect of nonlocal deformations in cosmology   could be a straightforward  explanation for cosmic acceleration  \cite{nonl1, nonl2,Jhingan:2008ym,Nojiri:2010wj}. 

Furthermore, the nonlocality induced by GR deformations could be important 
to  understand better  the transition from radiation to matter 
dominated  era if 
consistently constrained   with the 
observations.

As nonlocality is produced by first order quantum gravitational effects, it is expected that they would also occur in Teleparallel Gravity. Unlike the standard local classical dynamics,  the behavior of such nonlocal effects could  be very different in Teleparallel Gravity and GR, and they can be used to experimentally discriminate between these two theories. Therefore, it is interesting to study the nonlocal deformation of both GR  and Teleparallel Gravity.  Even thought the nonlocal deformation of GR has been extensively studied, the nonlocal deformation of TEGR has not been studied. Thus, in this paper, we will analyze a model of nonlocal Teleparallel Gravity. 

We will observe that at present, the non-local Teleparallel Gravity satisfies all the existing cosmological experimental constraints, and can explain  phenomena that are explained using nonlocal deformations of GR. However, as the nonlocal Teleparallel Gravity is fundamentally different from nonlocal deformation of GR, future experiments can be used to verify which of these theories is the correct theory of gravity. Thus,   the action for general relativity  $\mathcal{S}_{\rm GR}$,  can be corrected by a nonlocal terms $\mathcal{S}_{\rm SRNL}$ due to quantum corrections, and so the quantum corrected nonlocal GR can be written as 
 \cite{nonl1, nonl2}
\begin{eqnarray}
\mathcal{S}_1 = \mathcal{S}_{\rm GR} + \mathcal{S}_{\rm GRNL}.
\end{eqnarray}
Similarly,  the standard classical action of  TEGR $\mathcal{S}_{\rm TEGR}$ can be corrected by a nonlocal term due to quantum corrections $\mathcal{S}_{\rm TEGRNL}$, and so the quantum corrected nonlocal Teleparallel Gravity can be written as 
\begin{eqnarray}
\mathcal{S}_2 = \mathcal{S}_{\rm TEGR} + \mathcal{S}_{\rm TEGRNL}.
\end{eqnarray}
It is not possible to experimentally differentiate between  $\mathcal{S}_{\rm GR}$ and $\mathcal{S}_{\rm TEGR}$, but the quantum corrections to these theories $\mathcal{S}_{\rm GRNL}$ and $\mathcal{S}_{\rm TEGRNL}$ are very different. Thus, it is experimentally possible to discriminate between $\mathcal{S}_1$ and $\mathcal{S}_2$. It may be noted that like in nonlocal GR case, the nonlocal correction to Teleparallel Gravity is motivated by quantum gravitational effects, and it is not arbitrary added to the original action. 

It may be noted that nonlocal teleparallel formalism could be a better approach  to study quantum gravitational effects. This is due to the fact that TEGR does not require the Equivalence Principle to be formulated (see Chapter 9 in \cite{tele1}), and it has been argued that quantum effects can cause the  violation of the Equivalence Principle \cite{equi1}. 
Furthermore, a violation of the Equivalence Principle can be related to a violation of the Lorentz symmetry \cite{equi2}, and 
 and Lorentz symmetry is also expected to be break  at the UV scale in various approaches to quantum gravity, such as  
discrete spacetime \cite{Hooft}, spacetime foam \cite{Ellis},
 spin-network in loop quantum gravity (LQG)
\cite{Gambini}, non-commutative geometry
\cite{Carroll,FaizalMPLA},   ghost condensation
\cite{FaizalJPA} and  Horava-Lifshitz gravity \cite{h1, h2}. In teleparallel theories of gravity, there are two different approaches. The first one does not assume that the spin connection (which is related to inertial effects) is zero making all the quantities invariant under Lorentz transformations. This formalism was implemented firstly in modifications of teleparallel theories of gravity in $f(T)$ gravity in \cite{Krssak:2015oua}. The second approach is the one where a specific frame is chosen at the beginning of the theory, or in other words, where one chooses the spin connection equal to zero. When Einstein and later Weitzenb\"ock formulated the teleparallel equivalent of general relativity theory, they chose that formalism. This approach is sometimes called the ``pure tetrad" formalism or the ``Weitzenb\"ock gauge" teleparallel formalism. In this formalism, the torsion tensor does not transforms covariantly under local Lorentz transformations. Hence, the torsion scalar also is not invariant under local Lorentz transformations. In standard teleparallel gravity where just a linear combination of the scalar torsion is considered in the action $\mathcal{S}_{\rm TEGR}$, the theory becomes quasi-local Lorentz invariant, or invariant up to a boundary term. However, when one is considering modifications of teleparallel theories of gravity, such as $f(T)$ gravity or in our case nonlocal teleparallel gravity, the theory is no longer local Lorentz invariant. In that case, in terms of computations, one way to alleviate this issue is by introducing the so-called ``good tetrad" \cite{Tamanini:2012hg}. Mostly all the papers related to $f(T)$ gravity work with this formalism so that in this work, we will follow it (see \cite{ftrev}). Further, the lost of this invariance in  teleparallel theories might be an interesting behavior on quantum scales. For a detail analysis related to the covariance of teleparallel theories of gravity, see \cite{Golovnev:2017dox,Krssak:2015oua,Krssak:2017nlv,Bahamonde:2017wwk,Arcos:2005ec,Li:2010cg}\\

In this paper, we will study a nonlocal deformation of Teleparallel Gravity, and the nonlocal  cosmological solutions obtained from such a deformed theory.  Furthermore,  we propose  a way to experimentally discriminate  Teleparallel Gravity from GR at quantum scales. 
The paper is organized as follows. In Sec.II, we discuss the action and the field equations of Nonlocal Teleparallel Gravity. Observational constrains coming from cosmology are given in Sec.III. These constraints result useful discriminate between nonlocal GR and Nonlocal Teleparallel Gravity. Conclusions are drawn in Sec.IV.
Appendix A is devoted to details in derivation of the field equations.

	\section{Nonlocal Teleparallel  Gravity }\label{sec:1}
    In this section, we will obtain  a nonlocal deformation of Teleparallel 
    Gravity. 
    Adopting  the formalism developed for  nonlocal deformations  
of GR \cite{nonl1,nonl2}, we can write a nonlocal deformation for    Teleparallel Gravity as 
\begin{eqnarray}
\mathcal{S}&=&\frac{1}{2\kappa}\int d^{4}x\, e(x) \, T(x)\Big[ f(\mathcal{G}[T](x))-1\Big]+\int d^{4}x\, e(x)\,L_{m}\\
&=&\mathcal{S}_{\rm TEGR}+\frac{1}{2\kappa}\int d^{4}x\, e(x) \, T(x)f\Big((\square^{-1}T)(x)\Big)+\int d^{4}x\, e(x)\,L_{m}\,, \label{1}
\end{eqnarray}	
	where $\kappa=8\pi G$, $T$ is the  torsion scalar ,  $e=\textrm{det}(e^{a}_{\mu})=\sqrt{-g}$,  $f$ is an arbitrary function which depends on the retarded Green function evaluated at the torsion scalar (quantum effects such as the Planck constant have been absorbed in the definition of this function), $L_{m}$ is any matter Lagrangian,  $\square \equiv \partial_{\rho}(e g^{\sigma\rho}\partial_{\sigma})/e$ is the 
  scalar-wave operator, and   $\mathcal{G}[f](x)$ is a nonlocal operator which can be written in terms of the  Green function $G(x,x')$ as 
\begin{eqnarray}
\mathcal{G}[f](x)&=&(\square^{-1}f)(x)=\int d^4x'\, e(x') f(x')G(x,x')\,.\label{G}
\end{eqnarray}
 Furthermore,  like the nonlocal corrections to the  GR,  these nonlocal corrections to the  Teleparallel Gravity are also motivated from quantum gravitational effects. We  note that,   as for nonlocal GR,  the  Green function is evaluated at the Ricci scalar $R$, in nonlocal Teleparallel  Gravity, the Green function is  evaluated at the torsion scalar $T$ (for the sake of simplicity, we write    $T(x)$ as $T$  and $e(x)$ as $e$). \\ 
It is worth noticing  that (unlike GR which produces the same equations of motion as the TEGR), the nonlocal deformation of GR   is different from the nonlocal deformation of Teleparallel Gravity. The latter comes from the fact that $R=-T+B$, where $B$ is a boundary term so that $\mathcal{S}_{GR}$ (which is constructed by $R$) and $\mathcal{S}_{TEGR}$(which is constructed by $T$) produces the same field equations. However, the nonlocal terms $\sqrt{-g}Rf_1(\Box^{-1}R)$
 and $eTf_2(\Box^{-1}T)$ coming from the nonlocal actions $\mathcal{S}_{\rm GRNL}$ and $\mathcal{S}_{\rm TEGRNL}$ will produce different field equations even for the case where $f_1=\Box^{-1}R$ and $f_2=\Box^{-1}T$. This happens since the boundary term $B$, which is the difference  between $T$ and $R$,  produces a contribution in the variational process in   nonlocal terms. This fact is in the same spirit as it was discussed in \cite{Bahamonde:2015zma,Bahamonde:2017wwk}, where it was shown that $f(R)$ and $f(T)$
 gravity (generalizations of $\mathcal{S}_{\rm GR}$ and $\mathcal{S}_{\rm TEGR}$ respectively), are different for this boundary term and the way to connect these two theories is to consider a more general action where the function depends on both the boundary term and the scalar torsion, the so-called $f(T,B)$ gravity (see also \cite{sebsal,Bahamonde:2016cul}). Moreover, the same happens when one considers more general theories like modified Gauss-Bonnet $f(R,G)$ gravity \cite{felicia} and teleparallel modified Gauss-Bonnet gravity $f(T,T_{G})$ \cite{manos} where  two boundary terms $f(T,B,T_{G},B_{G})$ needs to be taken into account  in order  to connect the two theories (for more details, see \cite{Bahamonde:2016kba}). Similarly, it is also possible to construct a general scalar tensor theory by considering non-minimally couplings between the scalar field and both the scalar torsion and the boundary term (see \cite{Bahamonde:2015hza,Zubair:2016uhx}). By doing that, one can also recover other well-known scalar tensor theories such as quintessence or non-minimally coupled curvature-scalar field theory.  Exactly as in those cases, in principle, one can extend the action (\ref{1}) changing $eTf(\Box^{-1}T)$ by $e f_1(T,B)f_2(\Box^{-1}T,\Box^{-1}B)$  and hence we achieve a more general theory which can connect Nonlocal Teleparallel Gravity with nonlocal GR for the cases $f_{1}=-T+B$ and $f_2=-\Box^{-1}T+\Box^{-1}B$.  \\
Now by a variation with respect to the tetrad, we obtain 
	\begin{eqnarray}
	\delta \mathcal{S}=\delta\mathcal{S}_{\rm TEGR}+\frac{1}{2\kappa}\int \Big[  T f(\mathcal{G}[T])\delta e+e f(\mathcal{G}[T])\delta T+e\, T\delta f(\mathcal{G}[T]) \Big]d^{4}x+\int d^{4}x\delta(eL_{m})\,, 
	\end{eqnarray}
where
\begin{eqnarray}
	e f(\mathcal{G}[T]) \delta T &=& -4\Big[
	e(\partial_{\mu}f(\mathcal{G}[T]))S_{a}\,^{\mu\beta}+\partial_{\mu}(e S_{a}\,^{\mu\beta})f(\mathcal{G}[T])-ef(\mathcal{G}[T])T^{\sigma}\,_{\mu a}S_{\sigma}\,^{\beta\mu}
	\Big]\delta e^{a}_{\beta} \,,
	\label{deltaT}\\[1ex]
	T f(\mathcal{G}[T]) \delta e &=&  eTf(\mathcal{G}[T]) E_{a}^{\beta} \delta e^{a}_{\beta} 
	\label{deltae1} \,,\\
	eT\delta f(\mathcal{G}[T])&=&e\Big[T\mathcal{G}[Tf'(\mathcal{G})]E_{a}^{\beta}+\partial_{\mu}(\mathcal{G}[Tf'(\mathcal{G})])(\partial_{\nu}T)\Big(g^{\mu\nu}E_{a}^{\beta}-2g^{\beta(\mu}E_{a}^{\nu)}\Big)\Big]\delta e^{a}_{\beta}\nonumber\\
	&&+e\,\mathcal{G}[Tf'(\mathcal{G})]\delta T\,.\label{deltaf}
\end{eqnarray}
See Appendix \ref{app} for  details on the variation of the nonlocal term (\ref{deltaf}). It is worth noticing  that the  energy-momentum tensor is 
\begin{equation}
\Theta^{\beta}_{a}=e^{-1} [{\delta (eL_{m})}/{\delta e^{a}_{\beta}}]\,,
\end{equation}  
so,  the field equations for Nonlocal Teleparallel Gravity can be written as 
\begin{eqnarray}
	4\Big[
	S_{a}\,^{\mu\beta}\partial_{\mu}+\frac{1}{e}\partial_{\mu}(e S_{a}\,^{\mu\beta})-T^{\sigma}\,_{\mu a}S_{\sigma}\,^{\beta\mu}
	-T\,E_{a}^{\beta}\Big]\Big[f(\mathcal{G}[T])+\mathcal{G}[Tf'(\mathcal{G})]\Big]\nonumber\\
	-\frac{4}{e}\partial_{\mu}(e S_{a}\,^{\mu\beta})+4T^{\sigma}\,_{\mu a}S_{\sigma}\,^{\beta\mu}+TE_{a}^{\beta}-\partial_{\rho}\Big(\mathcal{G}[Tf'(\mathcal{G})]\Big)(\partial_{\sigma }T)\Big(g^{\sigma\rho}E_{a}^{\beta}-2g^{\beta(\rho}E_{a}^{\sigma)}\Big)=2\kappa \Theta_{a}^{\beta}\,.
\end{eqnarray}
We have obtained the field equations for the nonlocal deformation of Teleparallel Gravity, and now we will analyze a nonlocal cosmological solution coming from this nonlocal model of gravity.   \\
Let us  assume a Friedman-Lema\^itre-Robertson-Walker (FLRW) cosmology with the following tetrad in Euclidean coordinates
$
e^{a}_{\beta}=(1,a(t),a(t),a(t)), \,\label{tetrad}
$ and write the   FLRW metric as  $ds^2=dt^2-a(t)^2(dx^2+dy^2+dz^2)$ for a spatially flat spacetime. We    will also consider a   power-law cosmology,  such that $
a(t)=a_{0}t^{s}\,,
$
where   $s$ is a constant. Now using the nonlocal formalism, we can observe that from Eq.  (\ref{G}), we obtain 
\begin{eqnarray}
\mathcal{G}[T]&=&-\int_{t^{*}}^{t}\frac{dt'}{e(t')}\int_{t^{*}}^{t'}dt''e(t'')T(t'')\,,\\
&=&\frac{6 s^2 }{(1-3 s)^2}\left[1-\left(\frac{t}{t^{*}}\right)^{1-3 s}\right]-\frac{6 s^2 \log \left(\frac{t}{t^{*}}\right)}{3 s-1}\,.
\end{eqnarray}
\begin{figure}
 	\includegraphics[width=8.0cm]{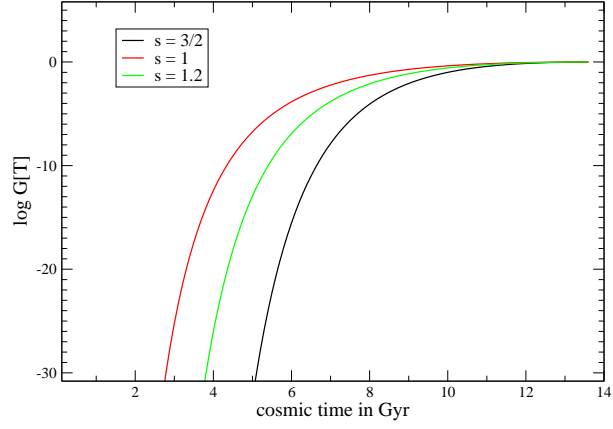}
 	\caption{\label{G_T} {\it{ Evolution of $\mathcal{G}[T]$ as a function of the cosmic time in Gyr for some values of $s$.}}}
\end{figure}
This can be used to analyze the effect of nonlocal deformation in Teleparallel Cosmology. From 
Fig. \ref{G_T} we can observe  the evolution of the function $\mathcal{G}[T]$ for the universe dominated by a certain form of matter ($s = 3/2$) and for
the universe dominated by a specific scalar field (s = $1$, $1.2$).
\section{Observational Constraints}
In this section, we will analyze some observational constraints for Nonlocal  Teleparallel Cosmology. As  discussed in \cite{Nojiri:2007uq}, to analyze the observational constraints, we  first 
  express  the nonlocal action in Eq.~(\ref{1}), in terms of   two scalar fields $\phi$ and $\xi$. In our case, we have
\begin{eqnarray}
\mathcal{S}&=&\frac{1}{2\kappa}\int d^{4}x\, e \Big[T(f(\phi)-1)-\partial_{\mu}\xi \partial^{\mu}\phi-\xi T\Big]+\int d^{4}x\, e\,L_{m}\,.\label{action2}
\end{eqnarray}
Now  we can assume $
\phi= \square^{-1}T\,, \, 
\square \xi= -f'(\phi)T\,, 
$ and  by varying this nonlocal action   with respect to the tetrads, we obtain 
\begin{eqnarray}
2(1-f(\phi)+\xi)\left[ e^{-1}\partial_\mu (e S_{a}{}^{\mu\beta})-E_{a}^{\lambda}T^{\rho}{}_{\mu\lambda}S_{\rho}{}^{\beta\mu}-\frac{1}{4}E^{\beta}_{a}T\right]\nonumber\\
-\frac{1}{2}\Big[(\partial^{\lambda}\xi)(\partial_{\lambda}\phi)E_{a}^{\beta}-(\partial^{\beta}\xi)(\partial_{a}\phi)-(\partial_{a}\xi)(\partial^{\beta}\phi)\Big] -2\partial_{\mu}(\xi-f(\phi))E^\rho_a S_{\rho}{}^{\mu\nu}=  \kappa\Theta^\beta_a\,. \label{2}
\end{eqnarray}
Thus, the field   equations can be written as 
\begin{eqnarray}
\label{H_function}
3H^2(1+\xi-f(\phi))&=&\frac{1}{2}\dot{\xi}\dot{\phi}+\kappa (\rho_{m} + \rho_{\Lambda})\,,\\
(1+\xi-f(\phi))(3H^2+2\dot{H})&=&-\frac{1}{2}\dot{\xi}\dot{\phi}+2H(\dot{\xi}-\dot{f}(\phi))-\kappa (p_{m} + p_{\Lambda})\,, 
\end{eqnarray}
and  the equations for the scalar fields can be written as 
\begin{eqnarray}
-6H^2 f'(\phi)+3H  \dot{\xi}+\ddot{\xi }&=&0\,,\\
3H\dot{\phi}+6H^2+\ddot{\phi}&=&0\,.\label{phi_function}
\end{eqnarray}

These  equations describe  a nonlocal model of Teleparallel Cosmology. We can  take into account  constraints on them from recent cosmological data.  
We will assume $f(\phi) = A \exp(n \phi)$, in order to test the dynamics of the model 
given by  the system (\ref{H_function})-(\ref{phi_function}). In order to constrain the free parameters of the model, 
we consider the following data sets:
\\

\noindent\textbf{ SNe Ia}: Type Ia supernovae (SNe Ia) have been  used to discover the current stage of accelerated
expansion of the universe. Hence, these observational data are a powerful tool for geometric tests. Here, 
let us adopt the latest ``joint light curves" (JLA) sample \cite{sn}, comprised of 740 type Ia supernovae 
in the redshift range  $0.01 \leq z \leq 1.30$.
\\

\noindent\textbf{BAO}: The baryon acoustic oscillations (BAO) is another important probe. 
We use the BAO measurements from the  Six  Degree  Field  Galaxy  Survey  (6dF) \cite{bao1}, 
the  Main  Galaxy  Sample  of  Data  Release 7  of  Sloan  Digital  Sky  Survey  (SDSS-MGS) \cite{bao2}, 
the  LOWZ  and  CMASS  galaxy  samples  of  the Baryon  Oscillation  Spectroscopic  Survey  (BOSS-LOWZ  and  BOSS-CMASS,  
respectively) \cite{bao3},  and the distribution of the LymanForest in BOSS (BOSS-Ly) \cite{bao4}.  
These data points are summarized in table I of \cite{baotot}.
\\

\noindent\textbf{CC+$H_0$}:  The cosmic  chronometers (CC) data set are another important data set. Here, we use the CC data set comprising
of 30 measurements spanned in the redshift range $0 < z < 2$, recently compiled in \cite{cc}. We also use
the recently measured new local value of Hubble constant given by $H_0= 73. 24 \pm 1.74$ km/s/Mpc.
\\

\begin{figure}
     	\includegraphics[width=10.0cm]{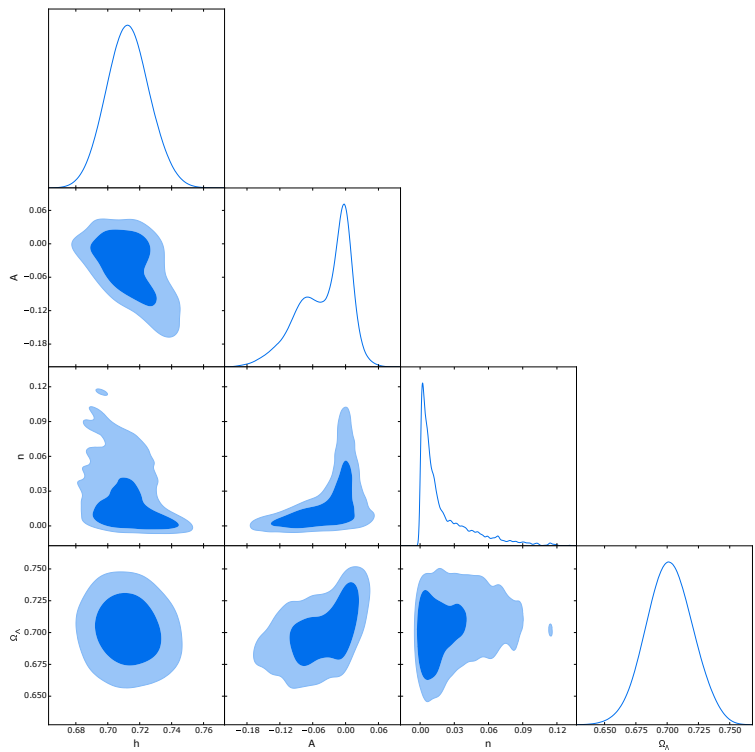}
     	\caption{\label{T_nonlocal} {\it{ One-dimensional marginalized distribution, and two-dimensional contours with 
     				68\% and 95\% confidence level for the free parameters of the model.}}}
\end{figure}

\begin{figure}
   	\includegraphics[width=8.0cm]{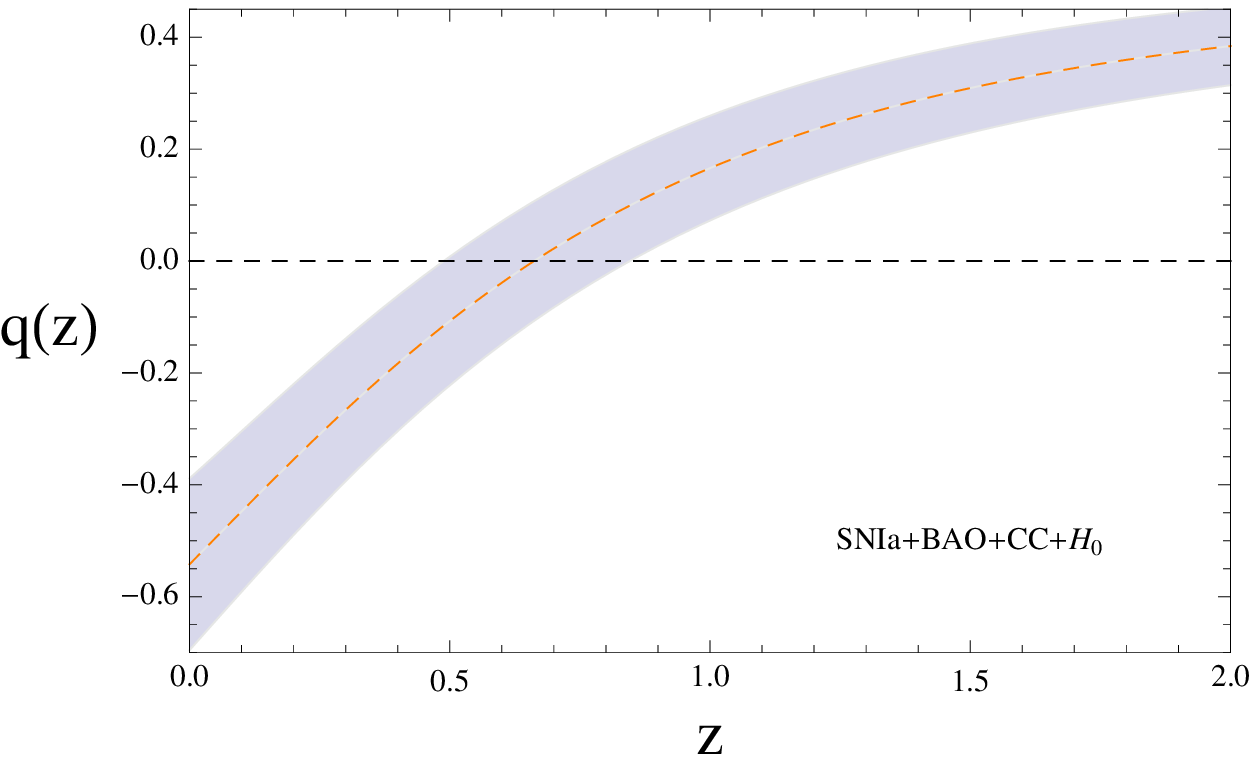}
   	\includegraphics[width=8.0cm]{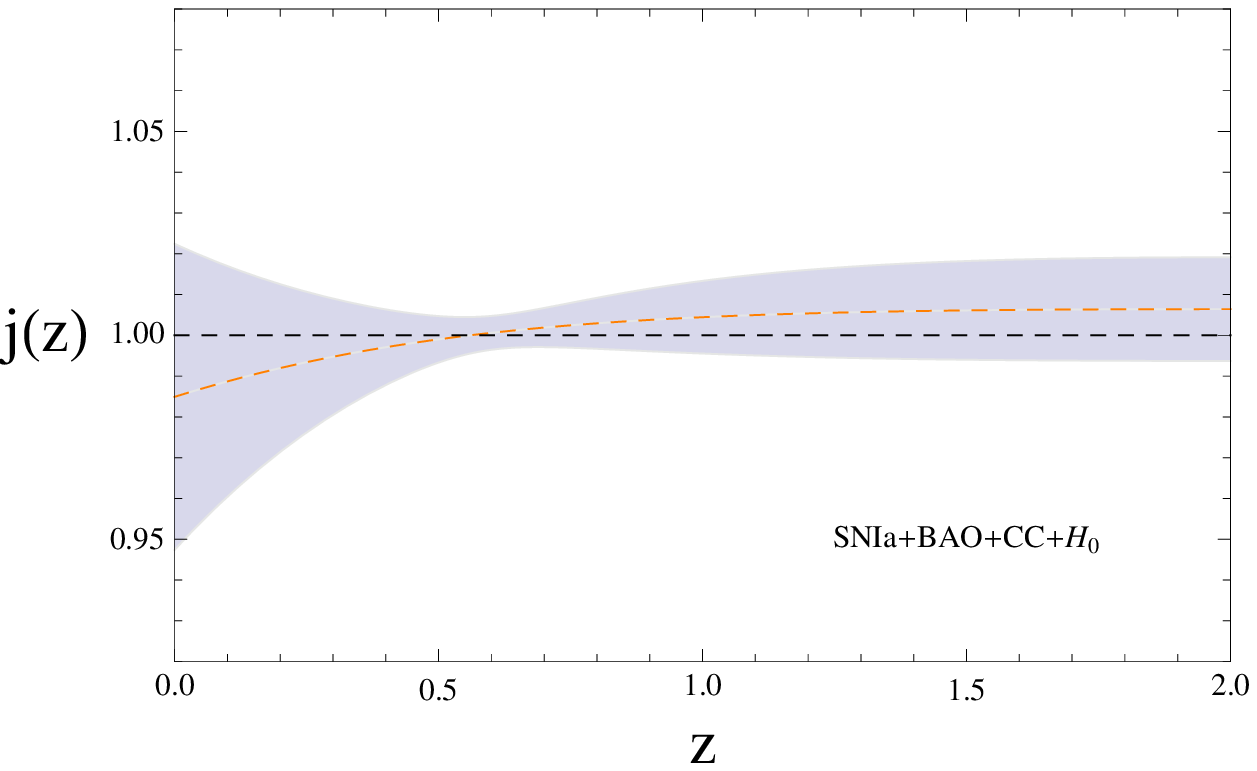}
   	\caption{\label{qz} {\it{ Reconstruction of the $q(z)$ (deceleration parameter) and $j(z)$ (jerk parameter) 
   	from $SNIa + BAO + CC + H_0$ data set at 1$\sigma$ CL.}}}
\end{figure}

We use the publicly available CLASS \cite{class} and Monte Python \cite{monte} codes for the model 
under consideration in orden to constrain the free parameters of this nonlocal cosmological model using SNe Ia + BAO + CC + $H_0$. 
We  used the Metropolis Hastings algorithm with uniform prior on the model parameters. In our analysis, 
we considered $\ddot{\phi} \ll \dot{\phi}$, $\ddot{\xi} \ll \dot{\xi}$. Figure \ref{T_nonlocal} 
shows the parametric space for $A$, $n$, $H_0$, and $\Omega_{\Lambda}$, 
at 1$\sigma$ and 2$\sigma$ confidence levels (CL) from the joint analysis SNIa + BAO + CC + $H_0$. 
We have observed at 1$\sigma$ CL the following constraints : $A=-0.009713_{-0.021}^{+0.017}$, $n=0.02086_{-0.0208}^{+0.0013}$, 
$h = 0.7127_{-0.015}^{+0.013}$ km/s/Mpc, $\Omega_{\Lambda }=0.7018_{-0.02}^{+0.018}$, and  $\Omega_{m0}=0.2981_{-0.018}^{+0.02}$, 
with $\chi^2_{min}=707.4$. We can note that the constraints are closed to the $\Lambda$CDM model, without any 
evidence for nonlocal effects in the present analysis, which here are characterized by the parameters $A$ and $n$.
In order to investigate kinematic effects, Figure \ref{qz} shows the deceleration and jerk parameters as a function of the redshift.
We consider the standard error propagation using the best fit values from SNIa + BAO + CC + $H_0$ in the reconstruction (gray region) 
of both parameters. On the left panel we have $q(z)$, where the transition from decelerated to accelerated phase occurs at $z \sim 0.6$, 
with $q_0 = -0.54 \pm 0.15 $. As expected, we have $q \rightarrow 1/2$ for high redshift. 
The right panel shows the jerk parameter $j(z)$ obtained from the joint analysis, the dotted black line ($j = 1$) represents 
the $\Lambda$CDM model. In general, small deviations can be noted when nonlocal effects are introduced, but
such effects are close the dynamics of $\Lambda$CDM model.
\\

The free parameters 
of the Nonlocal Teleparallel Cosmology are strongly constraint by present cosmological data. 
Furthermore, since  nonlocal GR and Nonlocal Teleparallel Gravity are fundamentally different,
it is possible that future cosmological data can be used to test which of these two proposals is the correct theory of gravity. 
As these theories are fundamentally different,  experiments can 
 be performed  to distinguish each other. Here we propose  some  possible 
experimental tests  that can be pursued in the near future to know which is the correct 
theory of gravity.  \\
The first experiment that can be performed is based on the violation of the 
   Equivalence Principle, as this can only occur in Nonlocal Teleparallel Gravity. The  
 accuracy of weak Equivalence Principle has been    measured from the  
 acceleration of Beryllium and Titanium test bodies using a rotating
torsion balance \cite{29}. It has been found that for acceleration $a$, the accuracy is 
of the order $\Delta a /a \sim 1.8 \time 10^{-13}$.
The accuracy is increased to $\Delta a /a \sim 2 \time 10^{-17}$ using the SR-POEM project 
\cite{26}. It is possible to use  more accurate future  experiments to observe a 
 violation  of the  weak Equivalence Principle. As such, a violation 
would only occur in Nonlocal Teleparallel Gravity and  it can be used as a experimental 
test to know which of these theories is the correct theory of nature. 
\\
We can also test these theories  by 
performing experiments using photon time delay and gravitational red shifts measured 
by high energy gamma rays. 
Both these nonlocal effects would produce 
different  photon time delays that have been observed 
by  measuring the round trip time of a  bounced radar beam off the surface
of Venus \cite{28}. This kind of  experiments, performed with more precision,  can be compared with  
 effects produced by the
nonlocal deformation of  both theories, and  any discrepancy between  results
can be used to discriminate between them. Similarly, gravitational red 
shift can be used to distinguish between the two theories. The   gravitational red shift derived by gamma rays of energy $14.4 \times 10^{-6} GeV$  has 
been measured  in the Pound-Snider experiment   \cite{27}, and it is 
  possible to perform similar experiments with higher energy gamma
rays with present day technology. Since Nonlocal Teleparallel Gravity and nonlocal GR 
predict different gravitational red sifting, such difference can be compared with these more 
accurate experiments.

\section{Conclusions}
Since GR and TEGR produce the same classical field equations, they cannot be differentiated  by using  classical experiments. However,  these theories are fundamentally different each other  and so  they have to produce  different quantum mechanical effects. According to this consideration, GR and Teleparallel Gravity could be   distinguished only at quantum level. Even though we do not have a fully developed theory of quantum gravity, there are several proposals in this direction. A universal prediction from almost all the approaches is the existence of  extended structures of spacetime geometry that  are expected to give rise to nonlocal deformations whose effects could be detectable from microscopic scales to cosmology. Thus, the nonlocal deformations of Teleparallel Gravity, just like the nonlocal deformations of GR,  are motivated by quantum gravitational effects. In principle, nonlocal cosmology from GR predicts a  different behavior with respect to  Nonlocal   Teleparallel Cosmology. Thus, the nonlocal deformations of these cosmological models  can be  matched with observational data. In analogy with a nonlocal deformation of GR,  we constructed a nonlocal deformation of Teleparallel Gravity. Starting from this, we derived nonlocal  cosmological solutions  and constrained  them using   data coming from SNeIa,  BAO, and  CC surveys. 
The main result of this paper is that  Nonlocal Teleparallel Gravity  is consistent with  present cosmological data and then cosmology, besides quantum experiments,
could be the ground on which discriminate the two approaches. 
As a general consideration,   nonlocal deformations for both GR  and Teleparallel 
Gravity are different, 
and the parameters of the field equations can be fixed, in principle,  by experiments.  
Here we  proposed also future experiments that can be performed to distinguish them 
from each other.

\begin{acknowledgments}
 SB is supported by the Comisi{\'o}n Nacional de Investigaci{\'o}n Cient{\'{\i}}fica y Tecnol{\'o}gica (Becas Chile Grant No.~72150066).
SC is supported in part by the INFN sezione di Napoli, {\it iniziative specifiche} TEONGRAV and QGSKY.
The  article is also based upon work from COST action CA15117 (CANTATA), 
supported by COST (European Cooperation in Science and Technology).  
\end{acknowledgments}

\appendix
\section{Derivation of the field equations}
\subsection{Variation of $f(\mathcal{G}[T])=f(\square^{-1}T)$}\label{app}
Let us consider the variation  of the action (\ref{1}) with respect to the tetrad fields. The term with the quantity $f(\square^{-1}T)$ is
\begin{eqnarray}
	eT\delta f(\mathcal{G}[T])=eT\delta f\Big(\square^{-1}T\Big)&=&eTf'(\mathcal{G})\Big(\frac{1}{\Box}\delta T-\frac{1}{\Box} (\delta \Box) \, \frac{1}{\Box}T\Big)\,\\
	&=&-e\Box^{-1}(Tf'(\mathcal{G}))\delta\Big(\frac{\partial_{\mu}eg^{\mu\nu}\partial_{\nu}}{e} \Big)\Box^{-1}T+e\Box^{-1}(Tf'(\mathcal{G}))\delta T\,.\label{ss}
\end{eqnarray}
We will not work out the second term on the right hand side since the variation of $F(e)\delta T$ is well-known for any function $F(e)$ which depends on the tetrad. Now, if we expand the first term,  we get
\begin{eqnarray}
	-e\Box^{-1}(Tf'(\mathcal{G}))\delta\Big(\frac{\partial_{\mu}eg^{\mu\nu}\partial_{\nu}}{e} \Big)\Box^{-1}T&=&(\Box^{-1}Tf'(\mathcal{G}))T\delta e-\Box^{-1}(Tf'(\mathcal{G}))\partial_{\mu}\delta(eg^{\mu\nu}\partial_{\nu})\Box^{-1}T\,,\\
	&=&T\mathcal{G}[Tf'(\mathcal{G})]\delta e+\partial_{\mu}(\mathcal{G}[Tf'(\mathcal{G})])(\partial_{\nu}T)\Big(g^{\mu\nu}\delta e+e\delta g^{\mu\nu}\Big)\,,
\end{eqnarray}
where we  used that $\Box \times \Box^{-1}T=T$ and we  neglected boundary terms. Now, if we take into account  that $\delta e=e E_{a}^{\beta}e_{\beta}^{a}$ and $\delta g^{\sigma\rho}=-(g^{\sigma\beta}E^{\rho}_{a}+g^{\rho\beta}E^{\sigma}_{a})\delta e_{\beta}^{a}$ we can expand the above term yielding
\begin{eqnarray}
	-e\Box^{-1}(Tf'(\mathcal{G}))\delta\Big(\frac{\partial_{\mu}eg^{\mu\nu}\partial_{\nu}}{e} \Big)\Box^{-1}T&=&e\Big[T\mathcal{G}[Tf'(\mathcal{G})]E_{a}^{\beta}+\partial_{\mu}(\mathcal{G}[Tf'(\mathcal{G})])(\partial_{\nu}T)\Big(g^{\mu\nu}E_{a}^{\beta}-2g^{\beta(\mu}E_{a}^{\nu)}\Big)\Big]\delta e^{a}_{\beta}\,.
\end{eqnarray}
Therefore, variations of the non-local term is 
\begin{eqnarray}
	eT\delta f(\mathcal{G}[T])&=&e\Big[T\mathcal{G}[Tf'(\mathcal{G})]E_{a}^{\beta}+\partial_{\mu}(\mathcal{G}[Tf'(\mathcal{G})])(\partial_{\nu}T)\Big(g^{\mu\nu}E_{a}^{\beta}-2g^{\beta(\mu}E_{a}^{\nu)}\Big)\Big]\delta e^{a}_{\beta}\nonumber\\
	&&+e\mathcal{G}[Tf'(\mathcal{G})]\delta T.
\end{eqnarray}


\begin{thebibliography}{99}



\bibitem{tele}Y. N. Obukhov and J. G. Pereira, Phys. Rev. D 67, 044016  (2003)

\bibitem{tele1} R. Aldrovandi,  J.G. Pereira, {\it Teleparallel Gravity: An Introduction.} Fund. Theor.  Phys., Vol. 173. Springer, Heidelberg (2013)

\bibitem{ft12}R. Ferraro and F. Fiorini, Phys. Rev. D 75, 084031  (2007)
\bibitem{ft14} G. R. Bengochea and R. Ferraro, Phys. Rev. D 79,  124019 (2009)

\bibitem{univ4} S. Das and E. C. Vagenas, Phys. Rev. Lett. 101, 221301 (2008)  
\bibitem{univ5}
I.  Pikovski, M. R. Vanner, M.  Aspelmeyer, M. Kim and  C.  Brukner, Nature Phys. 8, 393 (2012)  

\bibitem{univ1}  S. Masood, M. Faizal, Z. Zaz, A. F. Ali, J. Raza and M. B. Shah, 
Phys. Lett. B  763, 218 (2016)    

\bibitem{univ2}
M.  Faizal, A. F.  Ali and A. Nassar,  Phys. Lett. B 765 238 (2017)   

\bibitem{modesto1}
 L. Modesto, I.L. Shapiro,   Phys.Lett. B { 755}, 279  (2016).

\bibitem{modesto2}  L. Modesto,  Nucl.Phys. B { 909}, 584  (2016). 

\bibitem{st1}
G.  Calcagni and  L. Modesto, J. Phys. A: Math. Theor. 47, 355402 (2014)

\bibitem{st2}
G.  Calcagni and  G.  Nardelli, Phys. Rev. D82, 123518 (2010)

\bibitem{loop} S. A. Major and M. D. Seifert, Class. Quant. Grav. 19, 2211  (2002)  

\bibitem{z4} M. Maggiore, Phys. Lett. B304, 65 (1993)
\bibitem{z5}M. I. Park, Phys. Lett. B659, 698 (2008).


\bibitem{jm} L.  Modesto, J. W. Moffat and P. Nicolini,  Phys. Lett. B 695, 397 (2011)
\bibitem{jm12} C. Chicone and  B. Mashhoon,   Phys. Rev. D 87, 064015 (2013) 

\bibitem{Elizalde:1995tx}
E.~Elizalde and S.~D.~Odintsov,
Mod.\ Phys.\ Lett.\ A {\bf 10} (1995) 1821


\bibitem{nonl1}  S. Deser and  R. P. Woodard, Phys. Rev. Lett. 99, 111301 (2007)
\bibitem{nonl2}C. Deffayet   and  R. P. Woodard,
JCAP 0908, 023 (2009)

\bibitem{Jhingan:2008ym}
S.~Jhingan, S.~Nojiri, S.~D.~Odintsov, M.~Sami, I.~Thongkool and S.~Zerbini,
Phys.\ Lett.\ B {\bf 663} (2008) 424


\bibitem{Nojiri:2010wj}
S.~Nojiri and S.~D.~Odintsov,
Phys.\ Rept.\  {\bf 505} (2011) 59

\bibitem{numer} M. Kim, M. H. Rahat, M. Sayeb, L. Tan, R. P. Woodard and B. Xu, Phys. Rev. D 94, 104009 (2016)

 \bibitem{equi1}   L. Seveso and M. G. A. Paris, arXiv:1612.07331 [gr-qc]
 \bibitem{equi2} Z. Y. Wang, R. Y. Liu and X. Y.Wang, 
 Phys. Rev. Lett.   116,  151101 (2016)


\bibitem{Hooft} G. 't Hooft, Class. Quantum Gravit.  {13}, 1023 (1996)


\bibitem{Ellis} G. Amelino-Camelia, J. R. Ellis, N. Mavromatos, D. V.
Nanopoulos and S. Sarkar, Nature  {393}, 763 (1998)

\bibitem{Gambini} R. Gambini and J. Pullin, Phys. Rev. D  {59}, 124021
(1999)


\bibitem{FaizalMPLA} M. Faizal, Mod. Phys. Lett. A  {27}, 1250075
(2012)

\bibitem{Carroll} S.~M. Carroll, J.~A. Harvey, V.~A. Kostelecky, C.~D. Lane
and T.~Okamoto, Phys. Rev. Lett.  {87}, 141601 (2001)

\bibitem{FaizalJPA} M. Faizal, J. Phys. A  {44}, 402001 (2011)


\bibitem{h1} P.~Horava, Phys. Rev. D  {79}, 084008 (2009)

\bibitem{h2} P.~Horava, Phys. Rev. Lett.  {102}, 161301 (2009).


\bibitem{Krssak:2015oua}
M.~Krssak and E.~N.~Saridakis,
Class.\ Quant.\ Grav.\  {\bf 33} (2016) no.11,  115009

\bibitem{Tamanini:2012hg}
N.~Tamanini and C.~G.~Boehmer,
Phys.\ Rev.\ D {\bf 86} (2012) 044009


\bibitem{ftrev}
Y. F. Cai, S. Capozziello, M. De Laurentis, E. N. Saridakis,
Rept. Prog. Phys. {  79}, 106901 (2016)

\bibitem{Golovnev:2017dox}
A.~Golovnev, T.~Koivisto and M.~Sandstad,
Class.\ Quant.\ Grav.\  {\bf 34} (2017) no.14,  145013


\bibitem{Bahamonde:2017wwk}
S.~Bahamonde, C.~G.~Boehmer and M.~Krssak,
arXiv:1706.04920 [gr-qc].
\bibitem{Krssak:2017nlv}
M.~Krssak,
arXiv:1705.01072 [gr-qc].



\bibitem{Arcos:2005ec} 
H.~I.~Arcos and J.~G.~Pereira,
Int.\ J.\ Mod.\ Phys.\ D {\bf 13}, 2193 (2004)
\bibitem{Li:2010cg} 
B.~Li, T.~P.~Sotiriou and J.~D.~Barrow,
Phys.\ Rev.\ D {\bf 83}, 064035 (2011)








\bibitem{Bahamonde:2015zma}
S.~Bahamonde, C.~G.~B\"ohmer and M.~Wright,
Phys.\ Rev.\ D {  92}, 104042  (2015)


\bibitem{sebsal}
S.~Bahamonde and S.~Capozziello,
Eur.\ Phys.\ J.\ C {\bf 77} (2017) no.2,  107

\bibitem{Bahamonde:2016cul}
S.~Bahamonde, M.~Zubair and G.~Abbas,
arXiv:1609.08373 [gr-qc].


\bibitem{felicia}
M. De Laurentis and  A. J.  Lopez-Revelles, 
Int.  Jou.  Geom.  Meth. Mod. Phys.  { 11},  1450082 (2014).

\bibitem{manos}
G. Kofinas and E. N. Saridakis, Phys. Rev. D
{ 90}, 084044 (2014)

\bibitem{Bahamonde:2016kba}
S.~Bahamonde and C.~G.~B\"ohmer,
Eur.\ Phys.\ J.\ C {76},578  (2016)   

\bibitem{Bahamonde:2015hza}
S.~Bahamonde and M.~Wright,
Phys.\ Rev.\ D {\bf 92} (2015) no.8,  084034
Erratum: [Phys.\ Rev.\ D {\bf 93} (2016) no.10,  109901]
\bibitem{Zubair:2016uhx}
M.~Zubair, S.~Bahamonde and M.~Jamil,
Eur.\ Phys.\ J.\ C {\bf 77} (2017) no.7,  472


\bibitem{Nojiri:2007uq}
S.~Nojiri and S.~D.~Odintsov,
Phys. Lett. B { 659}, 821 (2008)  

\bibitem{sn}    M. Betoule et al., (SDSS collaboration), Astron. Astrophys. 568, A22 (2014), arXiv:1401.4064.

\bibitem{bao1} F. Beutler, C. Blake, M. Colless, D. H. Jones, L. Staveley-Smith, L. Campbell, Q. Parker, W. Saunders, and F. 
Watson, MNRAS  {416}, 3017 (2011) 

\bibitem{bao2}  A. J. Ross, L. Samushia, C. Howlett, W. J. Percival, A. Burden, and M. Manera, 
MNRAS  {449}, 835 (2015) 

\bibitem{bao3} L. Anderson, et  al., MNRAS  { 441}, 24 (2014) 

\bibitem{bao4} A. Font-Ribera, et  al., JCAP  {5}, 27 (2014) 

\bibitem{baotot} R. C. Nunes, S. Pan, E. N. Saridakis, and E. M. C. Abreu, JCAP 1701 01 005 (2017), arXiv:1610.07518.

\bibitem{cc} M. Moresco, R. Jimenez, L. Verde, A. Cimatti, L. Pozzetti, C. Maraston, and D. Thomas, arXiv:1604.00183 [astro-ph.CO].


\bibitem{class} D. Blas, J. Lesgourgues, and T. Tram,
JCAP  {07}, 034 (2011) 

\bibitem{monte} B. Audren, J. Lesgourgues, K. Benabed and S. Prunet, JCAP {02}, 001 (2013) 


 \bibitem{29} S. Schlamminger, K. Y. Choi, T. A. Wagner, J. H. Gundlach and 
 E. G. Adelberger, Phys. Rev. Lett. 100, 041101 (2008) 
 \bibitem{26} R.  D.  Reasenberg,  B.  R.  Patla,  J.  D.  Phillips 
 and R. Thapa, Class. Quant. Grav. 29, 184013 (2012)

\bibitem{28}I. Shapiro,  et. al., Phys. Rev. Lett. 26, 1132 (1971)

\bibitem{27}R. Pound and J. Snider, Phys. Rev. 140, B788 (1965)

   
\end{thebibliography}
\end{document}